\documentclass{agujournal}

\journalname{Water Resource Research}

\begin{document}

%
%

\title{The Cosmic-Ray Neutron Rover -- Mobile Surveys of Field Soil Moisture and the Influence of Roads}

%
%

 \authors{%
		M.~Schr{\"o}n\affil{1,2}, 
		R.~Rosolem\affil{2,3}, 
		M.~K{\"o}hli\affil{1,4,5}, 
		L.~Piussi\affil{6}, 
		I.~Schr{\"o}ter\affil{1}, 
		J.~Iwema\affil{2}, 
		S.~K{\"o}gler\affil{1}, 
		S.~E.~Oswald\affil{7}, 
		U.~Wollschl{\"a}ger\affil{8}, 
		L.~Samaniego\affil{9}, 
		P.~Dietrich\affil{1}, 
		S.~Zacharias\affil{1}, 
	} 

		\affiliation{1}{Dep. Monitoring and Exploration Technologies, Helmholtz Centre for Environmental Research - UFZ, Germany}
		\affiliation{2}{Faculty of Engineering, University of Bristol, England}
		\affiliation{3}{Cabot Institute, University of Bristol, England}
		\affiliation{4}{Physikalisches Institut, Heidelberg University, Germany}
		\affiliation{5}{Physikalisches Institut, University of Bonn, Germany}
		\affiliation{6}{Faculty of Science and Technology, Free University of Bolzano-Bozen, Italy}
		\affiliation{7}{Institute of Earth and Environmental Science, University of Potsdam, Germany}
		\affiliation{8}{Helmholtz Centre for Environmental Research - UFZ, Theodor-Lieser-Str. 4, 06120 Halle, Germany}
		\affiliation{9}{Dep. Computational Hydrosystems, Helmholtz Centre for Environmental Research - UFZ, Germany}
	

\correspondingauthor{Martin~Schr{\"o}n}{martin.schroen@ufz.de}


\begin{keypoints}
	\item road effect
	\item field-scale soil moisture
	\item cosmic-ray neutrons
	\item mobile survey
	\item COSMOS rover
\end{keypoints}

%
%

\begin{abstract}
Measurements of root-zone soil moisture across spatial scales of tens to thousands of meters have been a challenge for many decades. The mobile application of Cosmic-Ray Neutron Sensing (CRNS) is a promising approach to measure field soil moisture non-invasively by surveying large regions with a ground-based vehicle. Recently, concerns have been raised about a potentially biasing influence of local structures and roads. 
We employed neutron transport simulations and dedicated experiments to quantify the influence of different road types on the CRNS measurement.  We found that the presence of roads introduces a bias in the CRNS estimation of field soil moisture compared to non-road scenarios. However, this effect becomes insignificant at distances beyond a few meters from the road. Measurements from the road could overestimate the field value by up to \(40\,\%\) depending on road material, width, and the surrounding field water content.
The bias could be successfully removed with an analytical correction function that accounts for these parameters. Additionally, an empirical approach is proposed that can be used on-the-fly without prior knowledge of field soil moisture. Tests at different study sites demonstrated good agreement between road-effect corrected measurements and field soil moisture observations.  However, if knowledge about the road characteristics is missing, any measurements on the road could substantially reduce the accuracy of this method. 
Our results constitute a practical advancement of the mobile CRNS methodology, which is important for providing unbiased estimates of field-scale soil moisture to support applications in hydrology, remote sensing, and agriculture.
\end{abstract}

%
%

\section{Introduction}\label{introduction}

The monitoring of storage, movement, and quality of water at regional and global scales is of vital importance to practical applications such as agricultural production, water resources management, and predictions of flood, drought and climate change \citep{Seneviratne2010, Wood2011, Zink2016}. To study land surface processes, soil moisture information have to be included, averaged at a scale relevant and representative for the physical, chemical, and biological processes \citep{Entekhabi1999, Corwin2006, Schulz2006, Gentine2012, Vereecken2015}. The provision of parameters describing the critical processes at the landscape scale and capturing the natural heterogeneity of the soil-hydrological system at scales of 1 to 1000 m is one of the grand challenges in soil moisture monitoring \citep{Robinson2008, PetersLidard2017}.

Over the last 10--15 years, satellite-based Earth observation technologies made enormous progress. The potential of satellite-based remote sensing to map soil moisture dynamics at the catchment scale (\(\sim1000\,\mathrm{km^2}\)) has been demonstrated by numerous studies \citep{Kerr2007, Famiglietti2008, Wagner2009, WangQu2009, Liu2011, Ochsner2013}. While such data are widely used today to calibrate large-scale hydrological models up to the global scale, satellite-based remotely sensed soil moisture information are often not appropriate to reveal processes at the intermediate scale (up to 1000 m) \citep{Western2002}. The reasons for this are, e.g., the shallow measurement depth, disturbing influences of vegetation or surface roughness on the signal and resulting lacks in the data quality, and a too coarse spatial resolution \citep{Robinson2008}. The accuracy and precision of remotely sensed products is also not constant around the globe, which is less an issue for ground-based methods. Comparing the spatiotemporal coverage of remotely sensed soil moisture against spatiotemporal scales covered by local instruments (e.g.~TDR, gravimetry, EMI/ERT, gamma-rays, NMR), it becomes obvious that ``there is currently a gap in our ability to routinely measure at intermediate scales'' \citep{Robinson2008}.

The method of Cosmic-Ray Neutron Sensing (CRNS) for soil moisture estimation, introduced to the environmental science community by \citet{Zreda2008}, provides a much larger measurement footprint than any other ground-based local method. With a support volume in the order of \(10^4\) m\(^3\) (\(>100\,\)m radius, \(<0.8\,\)m depth, \citet{Koehli2015}), CRNS has a large potential to close the scale gap between point measurements of root-zone soil moisture and remotely sensed surface soil moisture \citep{Ochsner2013, Montzka2017}. The CRNS technology makes use of the extraordinary high sensitivity of cosmic-ray neutrons to hydrogen nuclei and measures the concentration of epithermal neutrons above the soil surface. Since its introduction, the CRNS technology has quickly established itself and is now used for soil moisture monitoring by many research groups operating worldwide \citep[e.g.,][]{Bogena2013, Franz2013, Peterson2016, Zhu2016, Schroen2017w}.

Pilot studies have shown the concept and potential of \emph{mobile} CRNS \citep{Desilets2010} using neutron detectors mounted on a ground-based vehicle (``rover''). The method is comparable to exploration missions with rovers on the Martian surface \citep{Jun2013}. With advances on understanding the CRNS method also for stationary probes, recent studies have more and more elaborated on direct applications of the so-called \emph{CRNS rover} \citep{Chrisman2013, McJannet2014, Dong2014, Franz2015, Avery2016}. While the ``classical'', stationary CRNS application enables to capture the hourly variability of soil moisture within a static footprint, the mobile application is intended to capture the spatial variability of soil moisture across larger areas or along larger transects. The CRNS rover uses the same detection principle as the stationary CRNS probes but deploys multiple and larger neutron detectors in order to achieve higher count rates at much shorter recording periods.

Agricultural fields and private land are often not accessible by vehicles. Hence, the CRNS rover is usually moved along a network of existing roads, streets, and pathways in a study region. This strategy is also practical when the rover is used to cover large areas at the regional scale in a short period of time. However, recent findings by \citet{Koehli2015} have shown that the CRNS detector is particularly sensitive to the first few meters around the sensor, which was later confirmed by calibration and validation campaigns of stationary CRNS probes \citep{Schroen2017w, Heidbuechel2016, Schattan2017}. This aspect is of even higher importance for the mobile application of CRNS. Following this argumentation, we hypothesize that the CRNS measurement is biased significantly when the moisture conditions present in the road differ substantially from the actual field of interest.

The effect of dry structures in the footprint was introduced for the first time by \citet{Franz2013} and was observed by \citet{Chrisman2013} on rover surveys through urban areas. \citet{Franz2015} sensed soil moisture of agricultural fields by roving on paved and gravel roads, and speculated that the road material could have introduced a dry bias to their measurements. The quantification of this effect is critical, not only for the advancement of the method, but also for its application to support agricultural irrigation management \citep{Franz2015} or to allow for large-scale soil moisture retrieval to support hydrological modeling \citep{Zink2016, Schroen2017diss} and evaluation of remote-sensing products \citep{Montzka2017}.

In the present study, we aim to evaluate and quantify the ``road effect'' by combining physical neutron transport modeling and dedicated field experiments. Based on theoretical investigations, we propose a universal correction function which is then tested and discussed in the light of ten rover campaigns in Central Germany and South England.

\section{Methods}\label{methods}

\subsection{The Cosmic-Ray Neutron Rover}\label{the-cosmic-ray-neutron-rover}

The method of cosmic-ray neutron sensing makes use of thermal neutron detectors filled with helium-3 or borontrifluorid \citep{Persons2011, Schroen2017u}. A surrounding shield of polyethylene prevents most thermal neutrons in the natural radiation environment from entering the detector, while it slows down incoming, epithermal neutrons to detectable, thermal energies \citep{Zreda2012, Andreasen2016}. The epithermal neutron density in air is mainly controlled by (1) the interaction of direct cosmic radiation with the ground, and (2) the number of hydrogen atoms in the environment \citep{Zreda2008, Koehli2015}. As hydrogen is an elemental part of the water molecule, the correlation between the epithermal neutron signal and surrounding water storages can be beneficial for the monitoring of the hydrological cycle. Figure \ref{fig:maps} shows a combination of the helium-3 detector system (white case, left), a small helium-3 unit (black case, middle), and four borontrifluorid tubes (right) which have been disassembled from stationary probes.

The mobile CRNS detectors can be mounted in the trunk of a car. As neutrons are almost exclusively sensitive to hydrogen, the metallic material of the car appears almost transparent. Additional plastic components and human presence can have the effect of a constant shielding factor, which is irrelevant for CRNS applications as only relative changes of neutrons are evaluated. Air temperature and humidity are recorded with sensors mounted externally to the car, because air conditions inside and outside can differ significantly. The neutron detector was set to integrate neutron counts over 1 minute. When in motion, this implicitly stretches the otherwise circular footprint to a patch elongated in the driving direction. In contrast, the GPS coordinates are read from a \texttt{Globalsat\ BR-355} sensor at the time of recording, so after the neutrons were integrated. To account for this artificial shift in post-processing mode, the UTM coordinates of each signal were back-projected to half of the distance covered within that minute. Driving speed was adapted on local structures and ranged from 15 to \(80\) meters per minute.
The neutron count rate \(N\) depends on environmental moisture conditions and on the detector volume used. The helium-3 detector system observed \(90\) to \(170\) counts per minute (cpm) and showed a similar count rate as the sum of four borontrifluorid tubes. Three consecutive measurements (i.e., 3 minutes) underwent a moving-average filter to account for the moving footprint and to reduce the relative statistical uncertainty, \(\sqrt{N}/N\), by a factor of \(\sqrt{3}\approx1.73\).

Besides near-surface water content, neutron radiation in the environment mainly depends on the incoming variation of cosmic rays, on the air mass above the sensor (and thus on altitude), and on water vapor in the air \citep{Schroen2015}. In this work, we have applied standard procedures to correct for these effects \citep{Zreda2012, Rosolem2013, Hawdon2014} in order to obtain a processed neutron count rate \(N\).

To convert the neutron count rate to gravimetric soil water equivalent, \(\theta_\text{grv}\), several approaches have been proposed in literature. \citet{Desilets2010} suggested a theoretical relation that has been applied successfully by the majority of CRNS studies in the past. \citet{McJannet2014} found that this approach performs also better for rover campaigns than the universal calibration function proposed by \citet{Franz2013u}, as the exact determination of soil and land-use data is the major obstacle to apply the latter. The standard approach from \citet{Desilets2010} is as follows:
\begin{linenomath*}
\begin{equation} \theta_\text{grv}(N,N_0) = \frac{a_0}{N/N_0-a_1}-a_2\,, \label{eq:desilets}\end{equation}
\end{linenomath*}
where parameters \(a_i=\{0.0808, 0.372, 0.115\}\) were determined using neutron physics simulations, and \(N_0\) is a (site-specific) calibration parameter. The latter is determined once for each dataset by comparing the CRNS soil moisture product with the actual soil moisture condition in the field. However, neutrons are sensitive to all occurances of hydrogen in the footprint, such as ponds, organic material, lattice water, plant water, and other dynamic contributors. Hence, the variable \(\theta_\text{grv}(N,N_0)=\theta_\text{sm}+\theta_\text{offset}\) denotes the sum of the soil water equivalent and an offset introduced by additional hydrogen pools. Furthermore, to compare CRNS products with other point sensors, the gravimetric water content is converted to volumetric water content, \(\theta_\text{vol}=\theta_\text{grv}\cdot\varrho_\text{bd}\), using soil bulk density information \(\varrho_\text{bd}\). In this work, we define
\begin{linenomath*}
\begin{equation} \theta(N)=\varrho_\text{bd}\,(\theta_\text{grv}(N,N_0)-\theta_\text{offset}) \label{eq:roversm}\end{equation}
\end{linenomath*}
as the CRNS soil moisture product, given in units of volumetric percent (\%) throughout this manuscript.

To account for spatially variable parameters of soil bulk density and land use throughout the study area, additional sources of data have been incorporated by recent studies \citep{Avery2016, Schroen2017diss, McJannet2017}. However, spatial information at the field scale (1--100 m) is often not available or come with significant uncertainty. This can be considered as a general handicap of the mobile CRNS method. In this work, we decided to apply the standard approach using spatially constant parameters, because (1) the selected study sites show sufficiently homogeneous soil and land-use conditions, and (2) the focus of the present study is to quantify the local effect of roads to the relative neutron signal, rather than the exact estimation of absolute soil moisture.

\subsection{Validation with point-scale measurements}\label{sec:weight}

Since the footprint of the CRNS signal covers an area of several hectares, comparison with point data is a challenge. To bridge this scale gap, \citet{Schroen2017w} developed a procedure to calculate a weighted average of point samples, based on their distance and depth to the neutron detector. The method uses an advanced spatial sensitivity function based on neutron transport simulations by \citet{Koehli2015}, and was successfully applied to calibration and validation datasets for stationary CRNS probes.

In our work presented here, we employed independent validation measurements of field soil moisture in the first 10 centimeters using occasional soil samples, and high frequency electromagnetic measurements with \texttt{Campbell\ TDR\ 100} and \texttt{Theta\ Probes}. The latter both instruments are standard approaches to determine near-surface soil water \citep{Roth1990}. The \texttt{Theta\ Probe} measures soil system impedance at \(100\,\)MHz, while the \texttt{TDR\ 100} evaluates pulse travel time in the GHz-range \citep[see also][]{Blonquist2005, Robinson2003, Vaz2013}.

In order to compare the point measurements with the CRNS soil moisture product, a weighted average of the point data is applied based on their individual distance \(r\) to the neutron detector (see illustration circle in Fig.~\ref{fig:maps}). Using eqs. \ref{eq:desilets}--\ref{eq:roversm}, the calibration parameter \(N_0\) can be determined from the neutron count rate \(N\) and the independently measured value for average field soil moisture, \(\langle\theta\rangle\).
The soil moisture products have been interpolated using an \emph{Ordinary Kriging} approach, as the chosen measurement density adequately represents typical spatial correlation lengths of soil moisture at our study sites. Furthermore, the Kriging approach supports the non-local nature of the epithermal neutron distribution in the air \citep{Franz2015, DesiletsZreda2013, Koehli2015}.

\subsection{Experimental setup}\label{sec:sites}

\subsubsection{Road types}\label{road-types}

Road moisture content is typically an uncertain quantity, only accessible by destructive sampling and lab analysis, or expensive geophysical exploration \citep{Saarenketo2000, Benedetto2012}. In the scope of the uncertainties involved in neutron sensing, e.g.~due to spatial heterogeneity of roads and surrounding land use, visual determination of the road material, guided by literature information, can allow for an adequate estimate of its elemental composition and thus, its soil water equivalent. \citet{Chrisman2013} analyzed several samples of stone/concrete and asphalt in Arizona and found their \emph{gravimetric} water equivalent to be \(1.52\,\%\) and \(5.10\,\%\), respectively. Following literature values for typical material densities from \(1.8\,\mathrm{g/cm^3}\) (sandy concrete) to \(2.4\,\mathrm{g/cm^3}\) (hot asphalt) \citep{Houben1994, Stroup2000}, the \emph{volumetric} water equivalent then is \(\approx3\,\%\) and \(\approx12\,\%\), respectively. As stone and asphalt are known as one of the most ``dry'' and ``wet'' road materials, respectively, we have estimated the moisture content of the various road types in our study sites within this range of extremes.

\subsubsection{Sch{\"a}fertal (Germany, experiment A)}\label{schuxe4fertal-germany-experiment-a}

The \emph{Sch{\"a}fertal} site is a headwater catchment in the \emph{Lower Harz Mountains} and one of the intensive monitoring sites in the \emph{TERENO Harz/Central German Lowland Observatory} (\(51^\circ\,39'\,\)N, \(11^\circ\,3'\,\)E) \citep{Zacharias2011, Wollschlaeger2016}. The catchment covers an area of 1.66 km\(^2\) and is predominantly under agricultural management. At the valley bottom, grassland surrounds the course of the creek \emph{Sch{\"a}ferbach}. Grassland is also present at the outlet of the catchment and a forest occupies a small area at the north-eastern end of the catchment outlet. Average climatology shows mean annual minimum and maximum temperatures of \(-1.8^\circ\)C and \(15.5^\circ\)C, respectively; and mean annual precipitation of 630 mm.
Average bulk density of the soil is \(\langle\varrho_\text{bd}\rangle=1.55\,\mathrm{g/cm^3}\) and water equivalent of additional hydrogen and organic pools have been approximated to be \(\langle\theta_\text{offset}\rangle=2.3\,\%\) in bare soil. For more information about the local hydrology, see \citet{Martini2015} and \citet{Schroeter2015}.

Within the Sch{\"a}fertal, \citet{Schroeter2015, Schroeter2017} performed regular TDR campaigns by foot using 94 locations in the whole catchment area. During several campaigns from 2014--2016, the CRNS rover accompanied their team. Shortly after harvest the fields were accessible with the car, such that the same locations could be sampled with the rover and the TDR team on the same campaign day.
On some days, however, the vehicle was only allowed to access the fields due to agricultural activities and seeded vegetation, such that CRNS measurements were taken on the sandy roads that were crossing the agricultural fields and the creek.

The road network consists mainly of three types: a paved major road between the hilltops and the urban area, sandy roads within the catchment, and pathways along the creek. The paved road has an average width of \(w\approx3.5\,\)m and consists of a very dry stone/concrete mixture that was estimated to contain \(\theta_\text{road}\approx4\,\%\) volumetric water content. The secondary roads are a mixture of stone, sand, and gravel, with \(\theta_\text{road}\approx6\,\%\) and width \(w\approx3\,\)m. The pathways of width \(w\approx3\,\)m contain mixed material from gravel, soil, and grass with an estimated average moisture content of \(\theta_\text{road}\approx10\,\%\).

Rover measurements have been performed using the helium-3 detector system at count rates of approximately \(90\) to \(170\,\)cpm depending on wetness conditions. The corresponding neutron count uncertainties of \(6\) to \(4\,\%\) propagated through eq.~\ref{eq:desilets} to absolute uncertainties in water equivalent, \(\Delta\theta_\text{grv}\), of \(10.0\) to \(0.9\) gravimetric percent, for wet to dry conditions, respectively.

\begin{figure}
\centering
\includegraphics{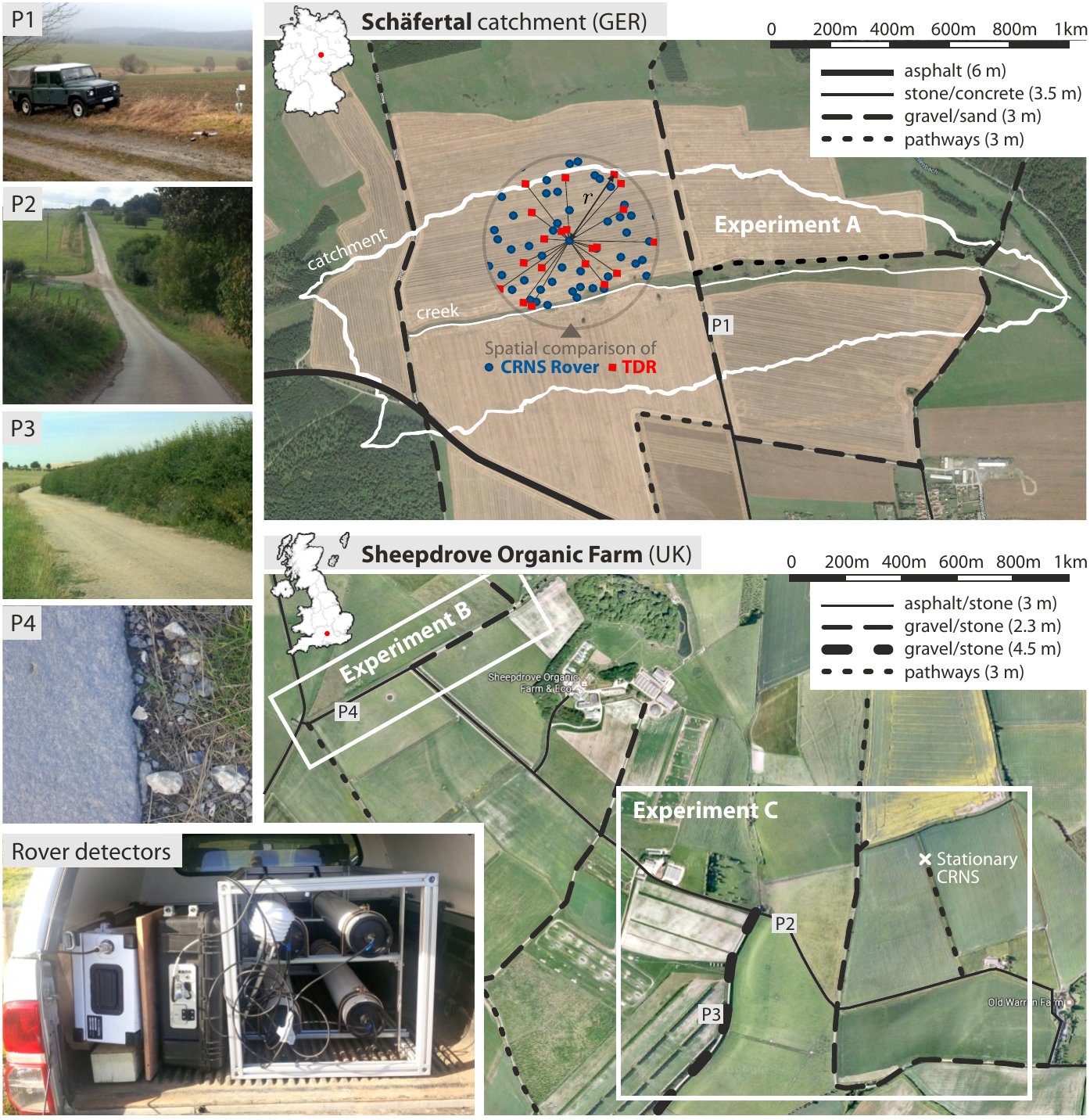}
\caption{The study sites \emph{Sch{\"a}fertal} (top right) and \emph{Sheepdrove Organic Farm} (bottom right). White borders indicate the areas of three different field experiments A, B, and C. Black lines indicate the type of road. The central circle with TDR points (red) and rover points (blue) illustrates the spatial calibration of the CRNS rover by comparing the large-scale neutron counts with point-scale soil moisture, using a weighted average of point samples based on their distance \(r\) to the rover. Pictures at certain spots: Sch{\"a}fertal gravel/sand road (P1), Sheepdrove valley (P2), Sheepdrove gravel/stone road, asphalt/stone road close-up (P4), and neutron detector tubes in the trunk of a car which is used as a CRNS rover.}\label{fig:maps}
\end{figure}

\subsubsection{Sheepdrove Organic Farm (England, experiments B and C)}\label{sheepdrove-organic-farm-england-experiments-b-and-c}

The \emph{Sheepdrove Organic Farm} is located on the \emph{West Berkshire Downs} in the Lambourn catchment in South England (\(51^\circ\,32'\,\)N, \(1^\circ\,29'\,\)W). The farm is located in a dry valley with elevations ranging from \(140\,\)m to \(200\,\)m above Ordnance Datum. The hydrogeology is characterized by a highly permeable white chalk aquifer with the groundwater table located tens of meters below the surface \citep{Evans2016}. Average climatology obtained from the Marlborough meteorological station (located 22 km to the south-west from the farm) shows mean annual minimum and maximum temperatures of \(5.4^\circ\)C and \(14.0^\circ\)C, respectively; and mean annual precipitation of 815 mm.
Soil information at the farm was collected at three sites with slightly different soil/vegetation characteristics between 2015 and 2017 \citep{Iwema2017}. The soil is generally loamy clay with many flints and pieces of chalk. Weathered chalk starts below the soil at about 25 to 40 centimeters depth. The average bulk density is \(\langle\varrho_\text{bd}\rangle=1.25\,\mathrm{g/cm^3}\) and water equivalent of additional hydrogen and organic pools have been determined to be \(\langle\theta_\text{offset}\rangle=4.3\,\%\) using soil sample analysis.

The road network consists of a paved major road (width \(w\approx3\,\)m) made of an asphalt/stone mixture with an estimated moisture content of \(\theta_\text{road}\approx11\,\%\). The main side roads are made of a gravel/stone mixture (\(\theta_\text{road}\approx7\,\%\)), most of which are \(w\approx2.3\,\)m wide while the southern road is \(w\approx4.5\,\)m wide. Many non-sealed tracks (\(w\approx3\,\)m) follow the borders between fields which partly consist of sand, grass, and organic material, such that their average moisture content was estimated to \(\theta_\text{road}\approx12\,\%\).

Rover measurements have been performed using the combination of the helium-3 detector system and the four borontrifluorid tubes at total count rates of approximately \(180\) to \(330\,\)cpm depending on wetness conditions. The corresponding neutron count uncertainties of \(4\) to \(3\,\%\) propagated through eq.~\ref{eq:desilets} to absolute uncertainties in water equivalent, \(\Delta\theta_\text{grv}\), of \(7.5\) to \(0.6\) gravimetric percent, for wet to dry conditions, respectively.

\subsection{Simulation of neutron interactions with road structures}\label{sec:theory}

Theoretical calculations of the CRNS footprint by \citet{Koehli2015} have shown that the radial sensitivity of a CRNS detector is strongly pronounced in the first few meters around the sensor \citep[see also][]{Schroen2017w}. Therefore, this work hypothesizes that there is an influence from the nearby road material to the neutron signal \(N\), which differs from the signal \(N_\text{field}\) measured above the soil if no roads were present. In this regard, we define the bias \(N/N_\text{field}\neq 1\) describing the relative deviation of measured neutrons \(N\) on the road from measurements on the field, \(N_\text{field}\), if the moisture contents of road and soil differ.

\begin{figure}
\centering
\includegraphics{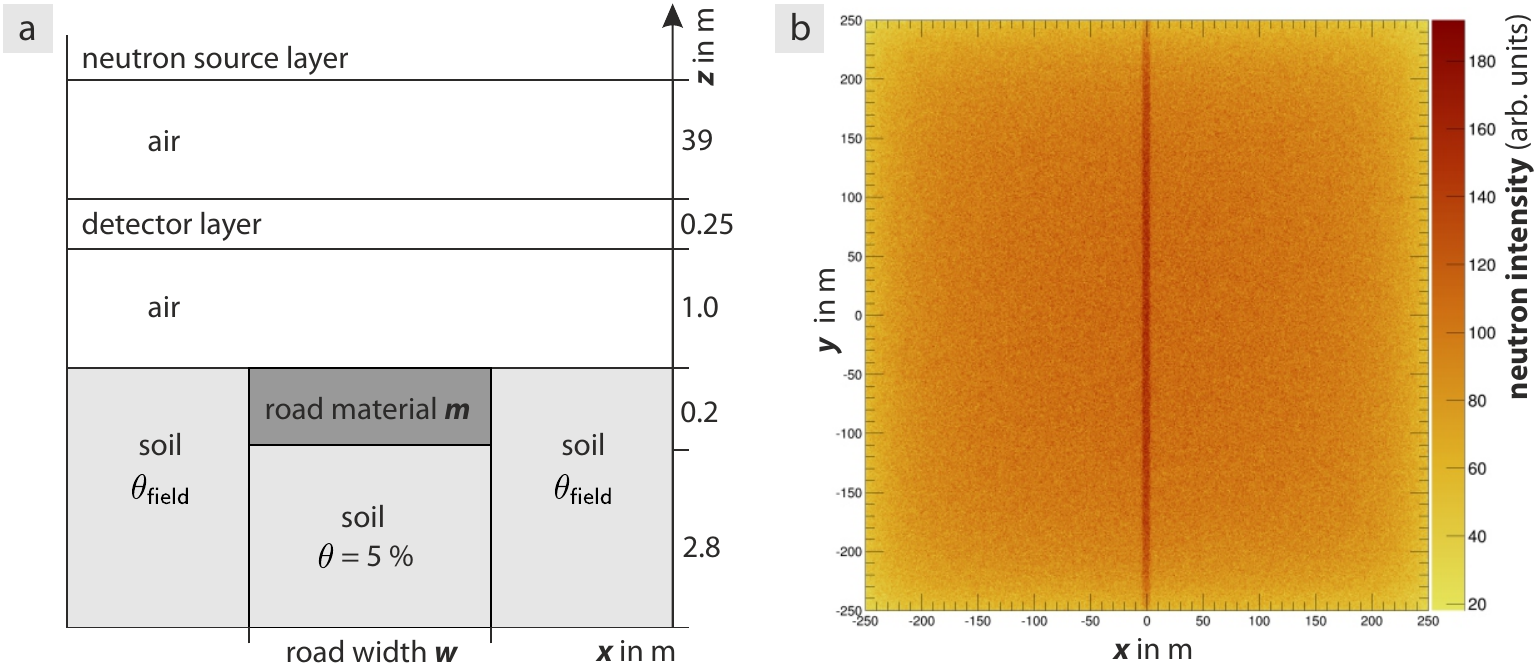}
\caption{(a) Schematic of the model setup used by the Monte-Carlo code \texttt{URANOS} to simulate the response of cosmic-ray neutrons to ground materials. (b) Exemplary \texttt{URANOS} model output showing a birds-eye view of the neutron density in the horizontal detector layer.}\label{fig:uranos}
\end{figure}

Many mobile surveys rely on road-only measurements of cosmic-ray neutrons, but we can expect that a potential road effect is larger for larger differences between road moisture and surrounding field water content. It is highly impractical to measure the corresponding bias rigorously, as it might depend also on the road material (see above), on field soil moisture, and on the distance to the road. We therefore employed the Monte Carlo technique using the neutron transport code \texttt{URANOS} (\citet{Koehli2015}, www.ufz.de/uranos) to simulate neutron response in a domain of 25 hectares which is crossed by a straight road geometry (see Fig.~\ref{fig:uranos}).

The road is modeled as a \(20\,\)cm deep layer of either stone or asphalt, while the soil below was set to \(5\,\%\) volumetric water content.
Following the compendium of material composition data \citep{McConn2011}, asphalt pavement is modeled as a mixture of O, H, C, and Si, at an effective density of \(\mathrm{2.58\,g/cm^3}\), which corresponds to a soil water equivalent of \(\approx12\,\%\). Stone/gravel is a mixture of Si, O, and Al, plus \(3\,\%\) volumetric water content at a total density of \(\mathrm{1.4\,g/cm^3}\) \citep{Koehli2015}.
The wetness of the surrounding soil has been set homogeneously to 10, 20, 30, and 40\(\,\%\) of volumetric water content. The neutron response to roads has been simulated for road widths of 3, 5, and 7\(\,\)m.

\clearpage

\section{Results \& Discussions}\label{results-discussions}

\subsection{Theoretical investigations}\label{theoretical-investigations}

The spatial Monte-Carlo simulations have been performed to study the interactions of cosmic-ray neutrons with roads of various widths, materials, and homogeneous field soil moisture conditions. The term ``relative road bias'' denotes the ratio of neutron intensity \(N\) detected in a road scenario (see Fig.~\ref{fig:uranos}) over neutron intensity \(N_\text{field}\) detected in a scenario of homogeneous soil moisture.

\begin{figure}[h]
\centering
\includegraphics{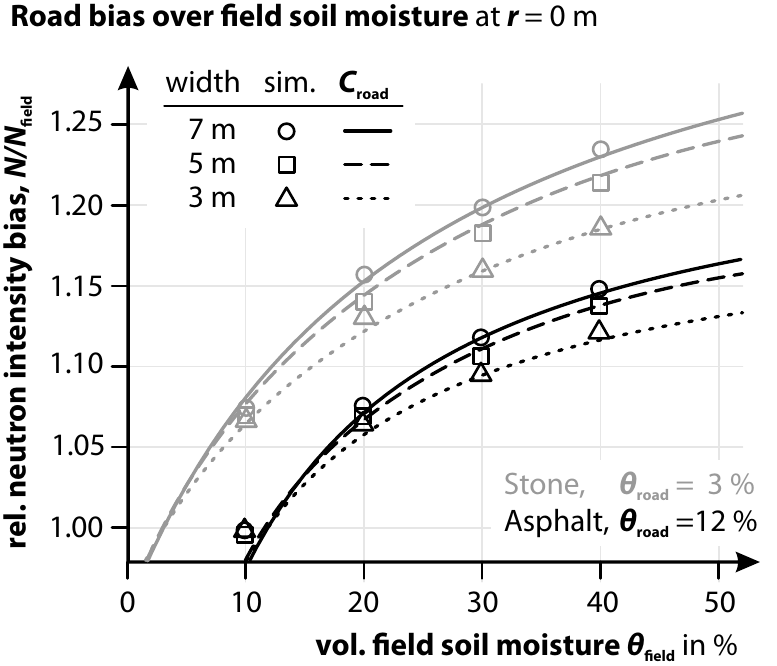}
\caption{\texttt{URANOS} simulations (symbols) and correction functions \(C_\text{road}(\theta_\text{field}, \theta_\text{road}, w, r)\) (lines) representing the neutron bias on roads of various widths through fields of different soil moisture. Shown for stone roads (grey) and asphalt roads (black).}\label{fig:croadsm}
\end{figure}

\begin{figure}
\centering
\includegraphics{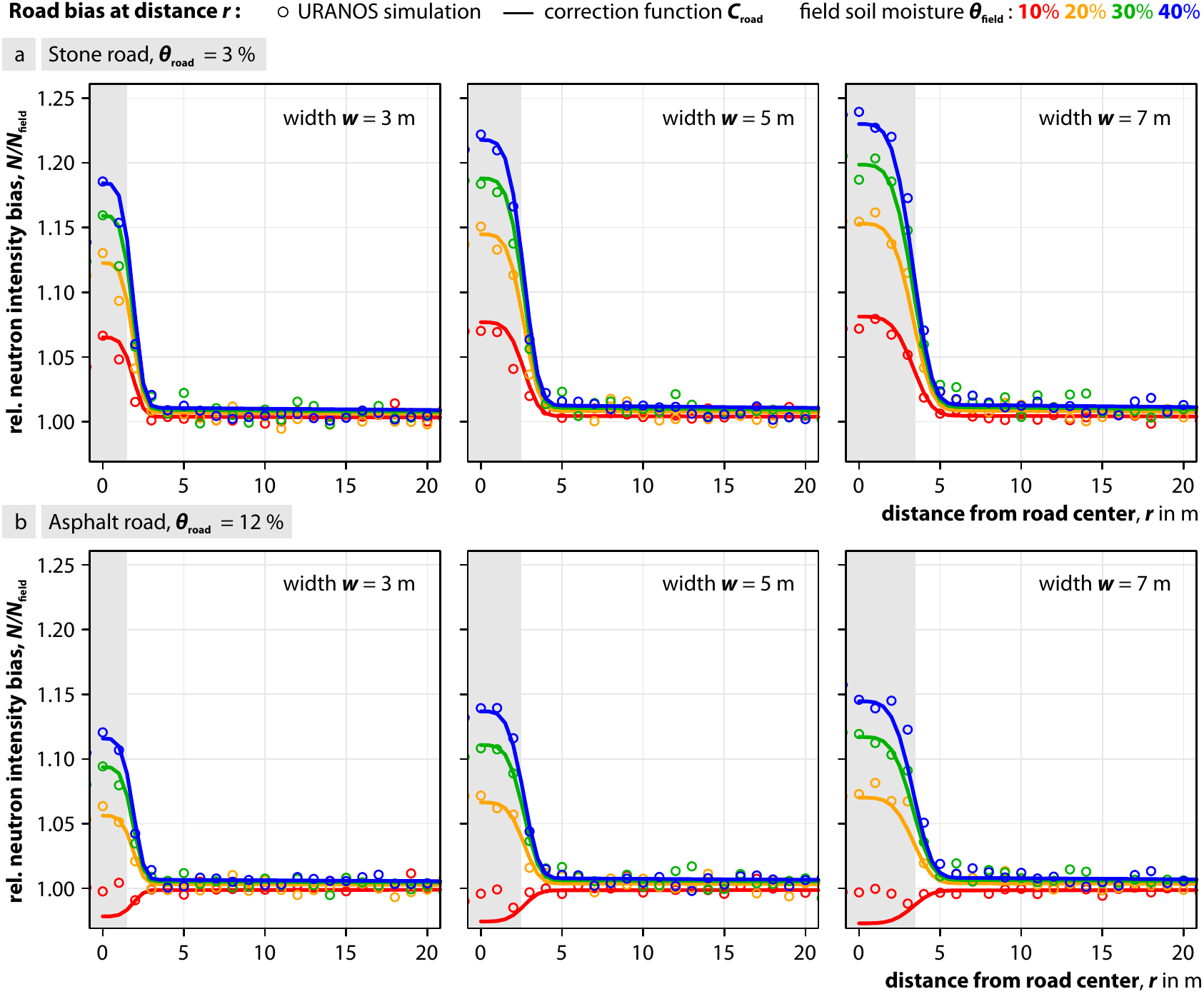}
\caption{\texttt{URANOS} simulations (circles) and correction functions \(C_\text{road}(\theta_\text{field}, \theta_\text{road}, w, r)\) (lines) representing the neutron bias at different distances \(r\) from the road center (\(r=0\)) for various road widths \(w\) (geometry shaded), field soil moisture (color), and (a) stone road material and (b) asphalt road material. Field conditions that are dryer than the road moisture (red in panel b) appear to be unresolved by the analytical approach.}\label{fig:croadr}
\end{figure}

Symbols in Fig.~\ref{fig:croadsm} show the simulated road bias for a detector placed at the center of the road. The bias increases for increasing field soil moisture, increasing road width, and decreasing road moisture. The quantity is particularly sensitive to the water equivalent of the pavement (\(\theta_\text{road}\)) and the soil (\(\theta_\text{field}\)).
Figure \ref{fig:croadr} plots the simulated road bias over distance from the road center, showing that the bias is a short-range effect that decreases a few meters away from the road, where almost no measurable effect can be expected beyond \(\approx10\,\)m distance. It is evident from these simulations that the road bias is higher the larger the difference between road moisture and surrounding soil moisture is and the wider the road.

We suggest to correct the observed neutron intensity with a correction factor \(C_\text{road}\), similar to the approaches used to correct for meteorological \citep{Hawdon2014, Schroen2015} and biomass effects \citep{Baatz2014}:
\begin{linenomath*}
\begin{equation}
N_\text{corr} = N / C_\text{road}\,,
\label{eq:croadapproach}\end{equation}
\end{linenomath*}
where the correction factor should be 1 for no-road conditions, plus a product of terms that depend on the characteristics of the road and field conditions. The shape of each term of the proposed correction function is based on physical reasoning as follows:

\begin{enumerate}
\def\labelenumi{\arabic{enumi}.}

\item
  The dependence on road width \(w\) is assumed to be a simple exponential, since the short-range dependency of neutron intensity on distance is exponential as shown by \citet{Koehli2015}.
\item
  The dependence on water content (\(\theta_\text{road}\) and \(\theta_\text{field}\)) is assumed to be hyperbolic, since the natural response of neutrons to soil water exhibits hyperbolic shape, as was derived from basic principles by \citet{Desilets2010} and \citet{Schroen2017diss}. This form (e.g., eq.~\ref{eq:desilets}) has been proven to be robust among all studies related to CRNS so far.
\item
  The dependence on distance \(r\) from the road center is assumed to be a sum of exponentials, since the combination of short- and long-range neutrons indicate this picture \citep[see][]{Koehli2015}. An additional polynom (\(w^a\, r^b\)) might be necessary to account for the plateau introduced by the road of a certain width \(w\).
\item
  Additionally, we demand that the total correction factor is 1 for road widths \(w=0\) and for similar moisture conditions, \(\theta_\text{road}=\theta_\text{field}\). The dependency on distance should be further normalized to 1 at the road center (\(r=0\)).
\end{enumerate}

The semi-analytical approach has been fitted to the \texttt{URANOS} simulation results. A minimum of 11 numerical parameters were required in order to adequately capture the most prominent features and dependencies of the simulated neutron response:

\begin{linenomath*}
\begin{equation}
C_\text{road}(\theta_\text{field}, \theta_\text{road}, w, r)
    = 1+F_1(w)\cdot F_2(\theta_\text{field},
    \theta_\text{road})\cdot F_3(r, w)\,,
\label{eq:croad}\end{equation}
\end{linenomath*}
where
\begin{linenomath*}
\begin{equation}
\begin{aligned}
F_1(w)
    &= p_0\,\big(1-e^{-p_1\,w}\big)\,,\\
F_2(\theta_\text{field}, \theta_\text{road})
    &= \big(\theta_\text{field}-\theta_\text{road}\big)\frac{p_2-p_3\,\theta_\text{road}}{\theta_\text{field}-p_4\,\theta_\text{road}+p_5}\,,\\
F_3(r, w)
    &= p_6\,e^{-p_7\,w^{-p_8}\,r^4}+p_9\,e^{-p_{10}\,r}\,.
\end{aligned}
\label{eq:croadp}\end{equation}
\end{linenomath*}
Parameters \(p_i\) of the geometry term \(F_1\), the moisture term \(F_2\), and the distance term \(F_3\) are given in Table~\ref{tbl:params}. Variables \(\theta_\text{field}\) and \(\theta_\text{road}\) are given in units of \(\mathrm{m^3/m^3}\), road width \(w\) and distance \(r\) are in units of m. The function is defined for road moisture values in the range of \(1\leq\theta_\text{road}\leq16\,\%\).

\clearpage

The function fits well to the simulation results for different distances \(r\) from the road center (Fig.~\ref{fig:croadr}), and for different \(\theta_\text{field}\), \(\theta_\text{road}\), and widths \(w\) (Fig.~\ref{fig:croadsm}). However, the analytical approach shows poorer performance for road widths of \(7\,\)m and beyond (not shown). The approach also overestimates the absolute bias when the field soil moisture becomes lower than the road moisture. These (rather unusual cases) should be avoided when the function is applied to roving datasets in the future. Since simulation results have indicated that the influence of slightly wetter road material is insignificant, a redefinition of the form \(F_2(\theta_\text{road}>\theta_\text{field})=1\) could be a sufficient approximation to these rare cases.

It is important to note that the moisture term \(F_2\) depends on prior knowledge of the field soil moisture \(\theta_\text{field}\).
The analysis of the field experiments in this work shows whether the moisture term can be replaced by a first-order approximation without prior knowledge.

\clearpage

\begin{table}
\caption{Parameters \(p_i\) of the parameter functions \(F_j\) describing the road correction factor \(C_\text{road}\) (eq.~\ref{eq:croadp}), namely the geometry term \(F_1\), the moisture term \(F_2\), the distance term \(F_3\), and the alternative moisture term \(F_{2'}\) that does not require prior information about field soil moisture (eq.~\ref{eq:croad2}).}
\label{tbl:params}
\begin{tabular}{lccccccccccc}
\hline
 & \(p_0\) & \(p_1\) & \(p_2\) & \(p_3\) & \(p_4\) & \(p_5\) & \(p_6\) & \(p_7\) & \(p_8\) & \(p_9\) & \(p_{10}\) \\\hline
\(F_1\) & 0.42 & 0.50 & & & & & & & & &\\
\(F_2\) & & & 1.11 & 4.11 & 1.78 & 0.30 & & & & &\\
\(F_{2'}\) & & & 1.06 & 4.00 & 0.16 & 0.39 & & & & &\\
\(F_3\) & & & & & & & 0.94 & 1.10 & 2.70 & 0.06 & 0.01\\\hline
\end{tabular}
\end{table}

\subsection{Experiment A: Estimating field soil moisture with TDR and the rover}\label{experiment-a-estimating-field-soil-moisture-with-tdr-and-the-rover}

Our first field experiment was designed in order to test the capabilities of the cosmic-ray neutron rover to capture small-scale patterns of soil moisture. During campaigns in the \emph{Sch{\"a}fertal}, the rover was moved across the fields over the course of four to six hours. At the rate of one data point per minute, the technology allowed to collect more than 200--400 points in the catchment, which is an adequate number to justify ordinary kriging within the \(1.66\,\mathrm{km^2}\) area.

\begin{figure}[h]
\centering
\includegraphics{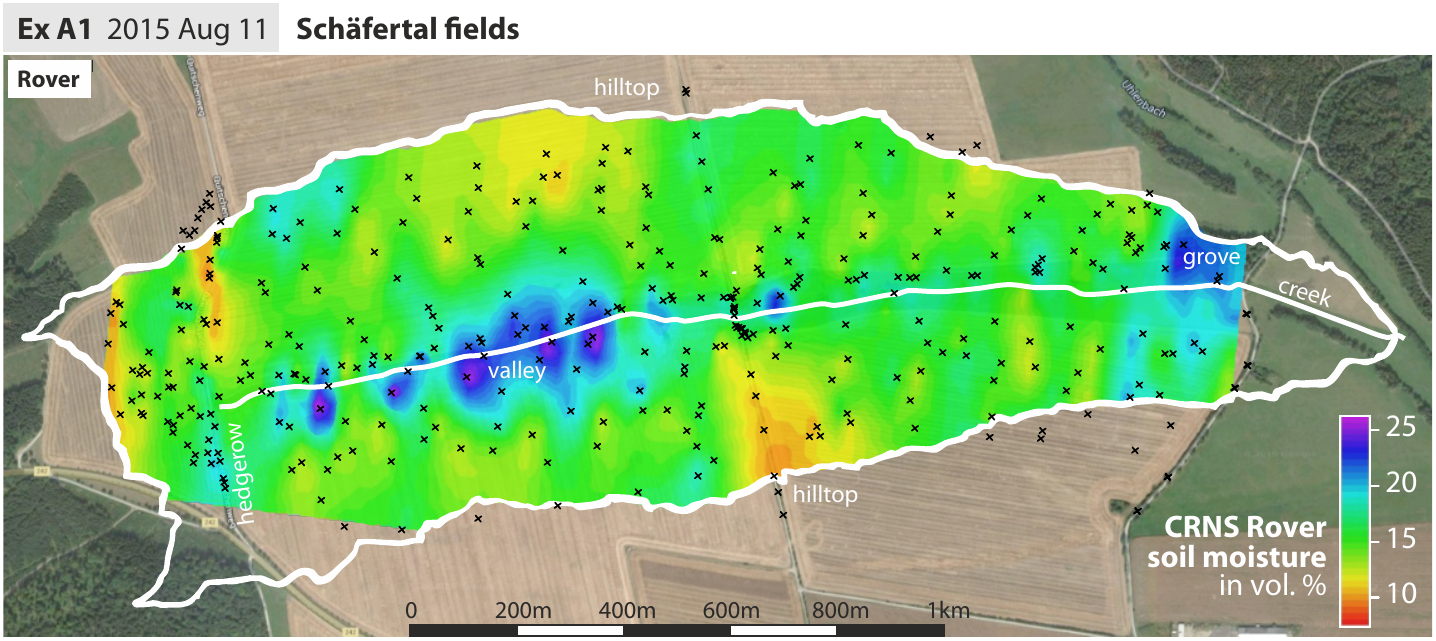}
\caption{Soil moisture estimation by the CRNS rover in the \emph{Sch{\"a}fertal} agricultural field. Data was interpolated from points (black cross) which represent the central location of the path travelled by the rover within the one minute acquisition time. Actual hydrological features like contact springs in the valley and dry hilltops are evident, but other influences of the grove, the hedgerow, and roads (see also Fig.~\ref{fig:maps}) may distort the derived soil moisture values, indicating challenges of the method.}\label{fig:schaefertal1}
\end{figure}

Figure \ref{fig:schaefertal1} shows the highly resolved CRNS soil moisture product which is able to reveal hydrological features in the catchment, such as dry hilltops, or contact springs in the valley near the creek due to shallow groundwater. Since the data were not corrected for biomass water, a probable influence of vegetation can be seen near the grove in the north-eastern part of the catchment, and possibly also near the hedgerow (south-western part). While the experiment focused on the agricultural areas of the harvested field and thus surveyed across the field and along its borders, a few roads were touched briefly at the southern and north-western hilltops, where the soil moisture appears to be slightly drier.

\begin{figure}
\centering
\includegraphics{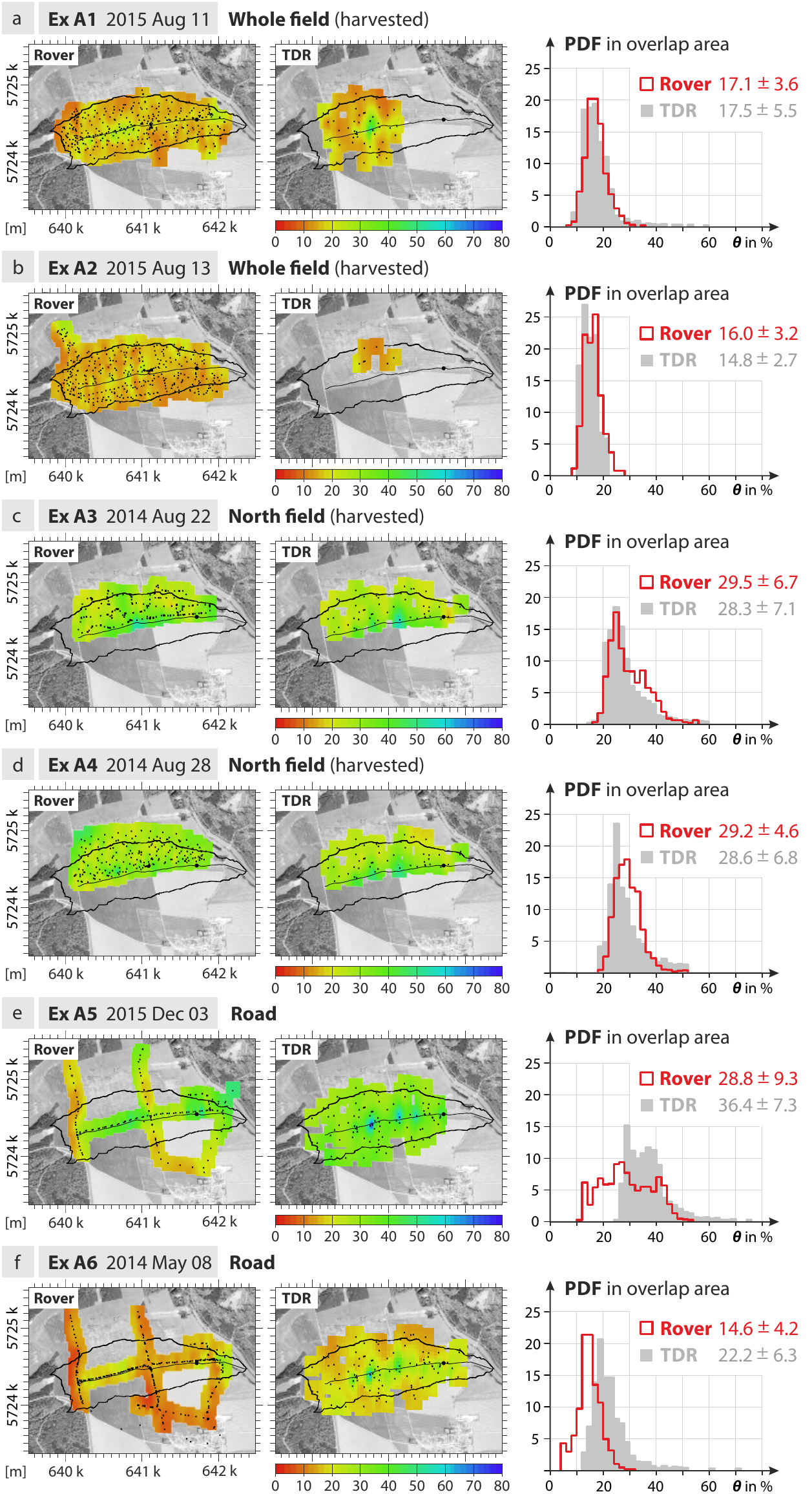}
\caption{Comparison of CRNS Rover and TDR campaigns in the \emph{Sch{\"a}fertal} using interpolated data, and the probability density functions (PDF) of their overlapping area.}\label{fig:schaefertal}
\end{figure}

Figure \ref{fig:schaefertal} summarizes results from this and other field surveys in the \emph{Sch{\"a}fertal} that were conducted together with a team using handheld TDR devices. Using 94 TDR samples and more than 300 rover points, it was possible to find the calibration factor \(N_0=10447\,\)cph (eq.~\ref{eq:desilets}) that explained all six sub-experiments in the catchment area.
In August 2015 (Fig.~\ref{fig:schaefertal}a,b), all the fields of the \emph{Sch{\"a}fertal} site were accessible with the car, however, TDR campaigns were incomplete due to technical issues. In the summer of 2014 (Fig.~\ref{fig:schaefertal}c,d), only the northern fields could be surveyed due to agricultural activities in the southern area.

For all of the first four campaign days, Fig.~\ref{fig:schaefertal}a--d, a good agreement between the rover and the TDR products in representing patterns and mean soil moisture in the \emph{Sch{\"a}fertal} was achieved. Besides the visual impression in columns 1 and 2, the probability density functions (third column) confirm emphatically that soil moisture patterns were well captured by both methods. The two approaches appear to show remarkable agreement, despite the fact that (i) the penetration depths of both methods were different (\(10\,\)cm for TDR versus 20--50\(\,\)cm for CRNS), (ii) TDR data was too sparse to achieve a comparable interpolation quality, and (iii) spatially constant parameters have been used for the calibration (\(N_0\), \(\varrho_\text{bd}\), \(\theta_\text{offset}\)).
In strong contrast, rover measurements at the last two survey days, Fig.~\ref{fig:schaefertal}e--f, show a poor agreement to field soil moisture measured by TDR. At those days, the CRNS rover had no access to the field and only crossed nearby roads and pathways. The corresponding impact on data interpretation is discussed in section~\ref{sec:A2}.

The field campaigns highlight characteristic hydrological features, e.g.~the mentioned contact springs near the creek, that are especially prominent during dry periods and which were identified also by other researchers using conventional measurement techniques \citep{Graeff2009, Schroeter2015}. The experiment shows that the rover can efficiently contribute to hydrological process understanding in small catchments, while the assumption of spatially constant parameters has shown to be acceptable for the almost homogeneous \emph{Sch{\"a}fertal} site.

\subsection{Experiment A: Taking the road effect into account}\label{sec:A2}

In May 2014 and December 2015, the fields were cultivated and the CRNS rover surveys were restricted to the roads. Those campaigns are shown in Fig.~\ref{fig:schaefertal}e,f, where the effect of the dry road is clearly visible in all panels.
This result indicates, that measurements only from the road are biased and therefore not representative for the field soil moisture. Under wet conditions, the probability density function (PDF) of soil moisture patterns becomes completely uncorrelated to the field conditions (Fig.~\ref{fig:schaefertal}e), while under dry conditions there seems to be a simple bias of the histogram towards the dry end (Fig.~\ref{fig:schaefertal}f).

\begin{figure}
\centering
\includegraphics{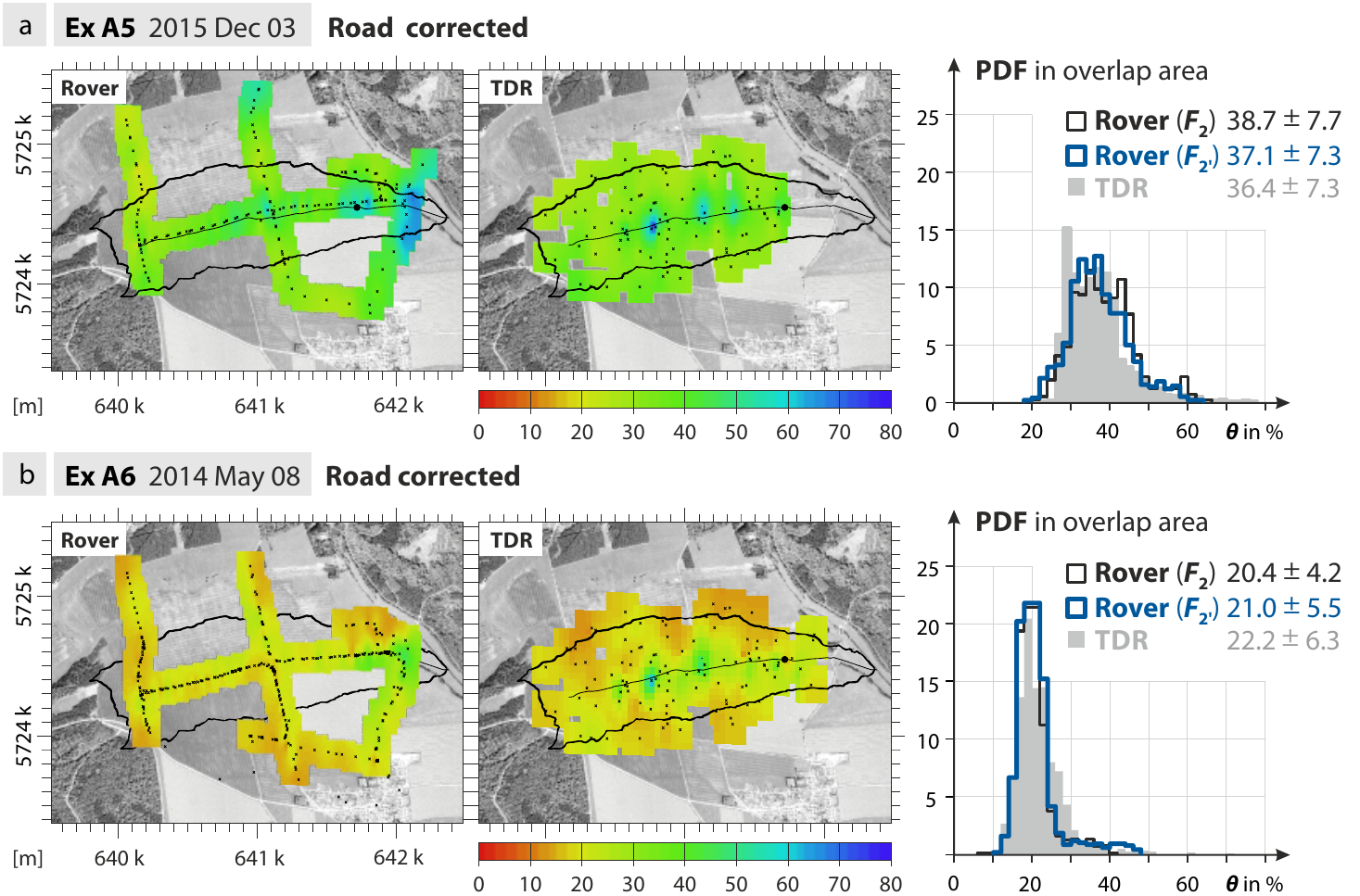}
\caption{Application of the road correction approach on the road-only surveys in the \emph{Sch{\"a}fertal} (compare Fig.~\ref{fig:schaefertal}e--f). Patterns of the rover (left) agree well with those from TDR (middle) in terms of the probability density function (right) in the overlap area of both interpolated grids, their mean, and standard deviation. The correction is tested with two approaches of the moisture term: (1) \(F_{2}(\theta_\text{field}=\langle\theta_\text{TDR}\rangle)\) (eq.~\ref{eq:croadp}) using the average of the TDR data (black line), and (2) \(F_{2'}(\theta(N))\) (eq.~\ref{eq:croad2}) using uncorrected neutron counts as a proxy (blue line). Kriging results using the former approach were almost identical to those using the latter approach, so that only the latter is shown in the left panel.}\label{fig:STcorrected}
\end{figure}

The presented road-effect correction approach promises to account for this behavior, as it scales with the difference between road and field moisture, using information of the different types of roads crossing the catchment (Fig.~\ref{fig:maps}). The correction function \(C_\text{road}\) was applied using prior knowledge about the mean field soil moisture (eq.~\ref{eq:croadp}). Using \(\theta_\text{field}=\langle\theta_\text{TDR}\rangle\) led to better agreement between the rover and the TDR data for both days as shown in Fig.~\ref{fig:STcorrected} (black histograms).

However, in most cases independent measurements of field soil moisture \(\theta_\text{field}\) are not available.
As an alternative, the first-order approximation of soil moisture, \(\theta(N)\), using the uncorrected neutron count rate \(N\), could be used as a proxy to estimate the bias due to the difference of soil moisture between road and field. An alternative analytical approach for the moisture term \(F_2\) (eq.~\ref{eq:croadp}) is proposed here that essentially accounts for the mismatch between \(\theta(N)\) and \(\theta_\text{field}\):
\begin{linenomath*}
\begin{equation}
F_{2'}(\theta(N), \theta_\text{road})
    \approx p_2-p_3\,\theta_\text{road}-\frac{p_4+\theta_\text{road}}{p_5+\theta(N)}\,,
\label{eq:croad2}\end{equation}
\end{linenomath*}
The updated empirical parameters \(p_\text{2--5}\) (Table~\ref{tbl:params}) have been determined based on the datasets of the \emph{Sch{\"a}fertal} and another, independent experiment in the context of an interdisciplinary research project which included rover measurements across different land-use types (Scale~X, see also \citet{Wolf2016}, data not shown here). Although the approach is empirical, the numerous tests through different sites and conditions indicate a robust potential. The corresponding probability distribution is indicated by a blue line in Fig.~\ref{fig:STcorrected}, showing that the two approaches led to almost identical results.

\subsection{Experiment B: Road influence at a distance}\label{experiment-b-road-influence-at-a-distance}

The experiments at the \emph{Sheepdrove Farm} aimed to compare the soil moisture patterns of the road and the field, by surveying both compartments with the rover and excluding the one or the other during the analysis. The general objective of these experiments was to clarify whether the road correction function is able to transfer the apparent soil moisture patterns seen from the road to values that were taken in the actual field.

The road network across the farm is an ideal location to test the road effect correction, due to its wide range of road materials (gravel to asphalt) and road widths (\(2.3\) to \(4.5\,\)m). To improve the accuracy of the rover measurements, the count rate was increased by combining multiple neutron detectors. Each of the rover datasets (experiments B and C) were compared to a stationary, well-calibrated CRNS probe and to occasional \texttt{Theta\ Probe} samples (not shown), in order to find a universal calibration parameter \(N_0=11300\,\)cph (see also eq.~\ref{eq:desilets}) for all datasets.

\begin{figure}
\centering
\includegraphics{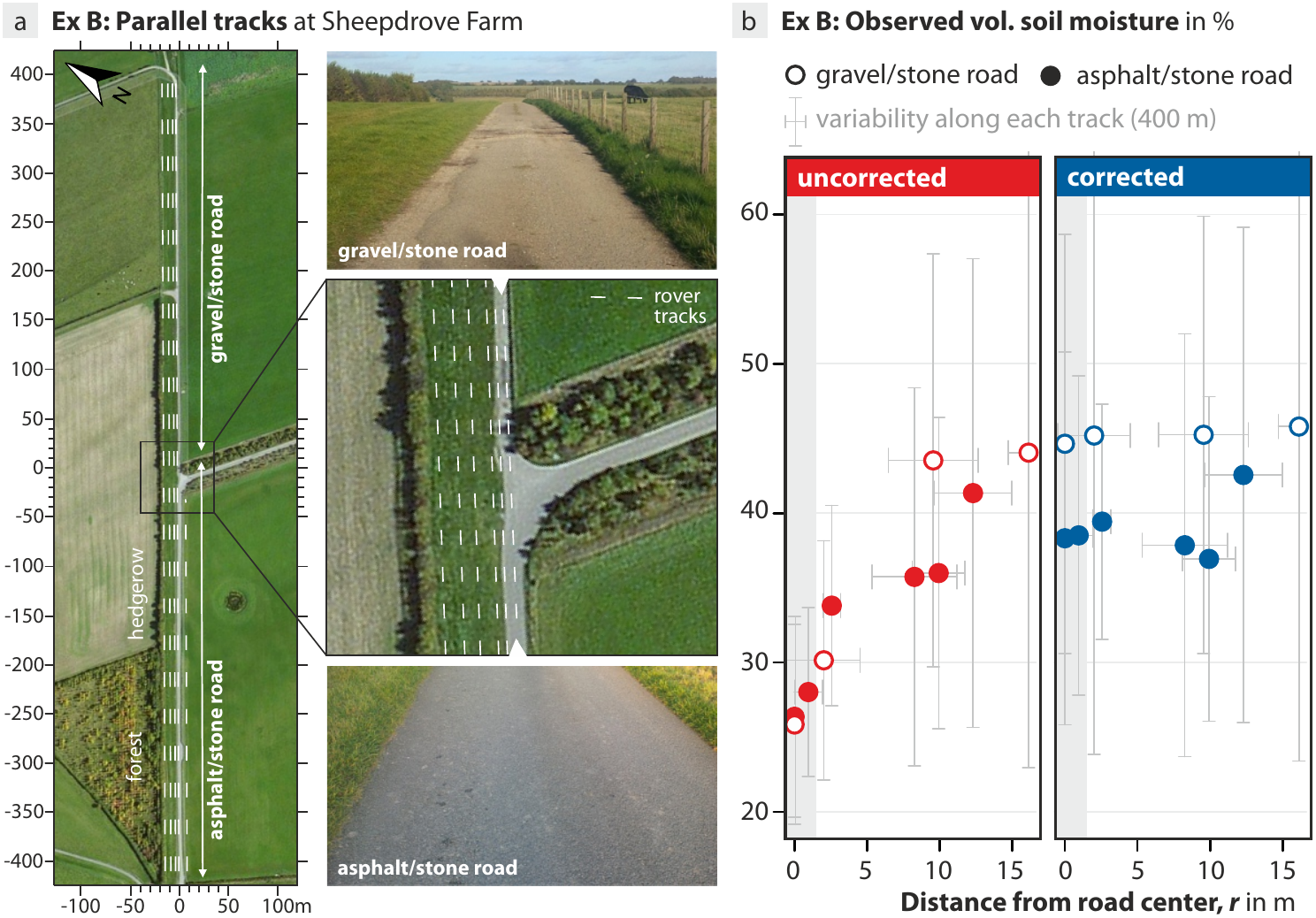}
\caption{Experiment B at the \emph{Sheepdrove Farm}. (a) Parallel tracks (dashed) at different distances from two roads of different materials that meet at the junction (\(x=0\), \(y=0\)). (b) The influence of the road decreases the apparent field soil moisture seen by the rover (left panel). Upon application of the road correction function, measurements converge to similar values for all distances (right panel) and reveal different soil moisture conditions for the nothern and southern fields. Error bars indicate the heterogeneity of water content along the whole track length of 400 m.}\label{fig:SFparallel}
\end{figure}

In order to rigorously test the theoretically predicted dependency of the road bias on the road moisture \(\theta_\text{road}\) and on the distance \(r\) to the road center, a dedicated experiment was performed at the north-west corner of the \emph{Sheepdrove Farm} (Fig.~\ref{fig:SFparallel}a). A gravel/stone road (north) and an asphalt/stone road (south) are aligned almost linearly and meet centrally at a junction. The road moisture reflects the mixture of present road materials and was estimated to be \(\approx11\,\%\) for the asphalt/stone mix, and \(\approx7\,\%\) for gravel/stone mix.

The rover measured neutrons along parallel lines in various distances from the road. For each track, the corresponding mean and standard deviation were calculated, which represent mainly the heterogeneity of soil and vegetation along each of the \(400\,\)m tracks. Fig.~\ref{fig:SFparallel}b shows how the influence of the road decreases the apparent field soil moisture as seen by the rover (left panel). Upon application of the road correction function, measurements converge to similar values for all distances (right panel) and reveal different soil moisture conditions for the nothern and southern fields. The apparent increase at \(r=12\,\)m is likely caused by hydrogen present in the hedgerow and the nearby grove. The overall result provides evidence that the analytical correction function properly represents the road bias at different distances and for different materials.

\subsection{Experiment C: Patterns across roads and fields}\label{experiment-c-patterns-across-roads-and-fields}

\begin{figure}
\centering
\includegraphics{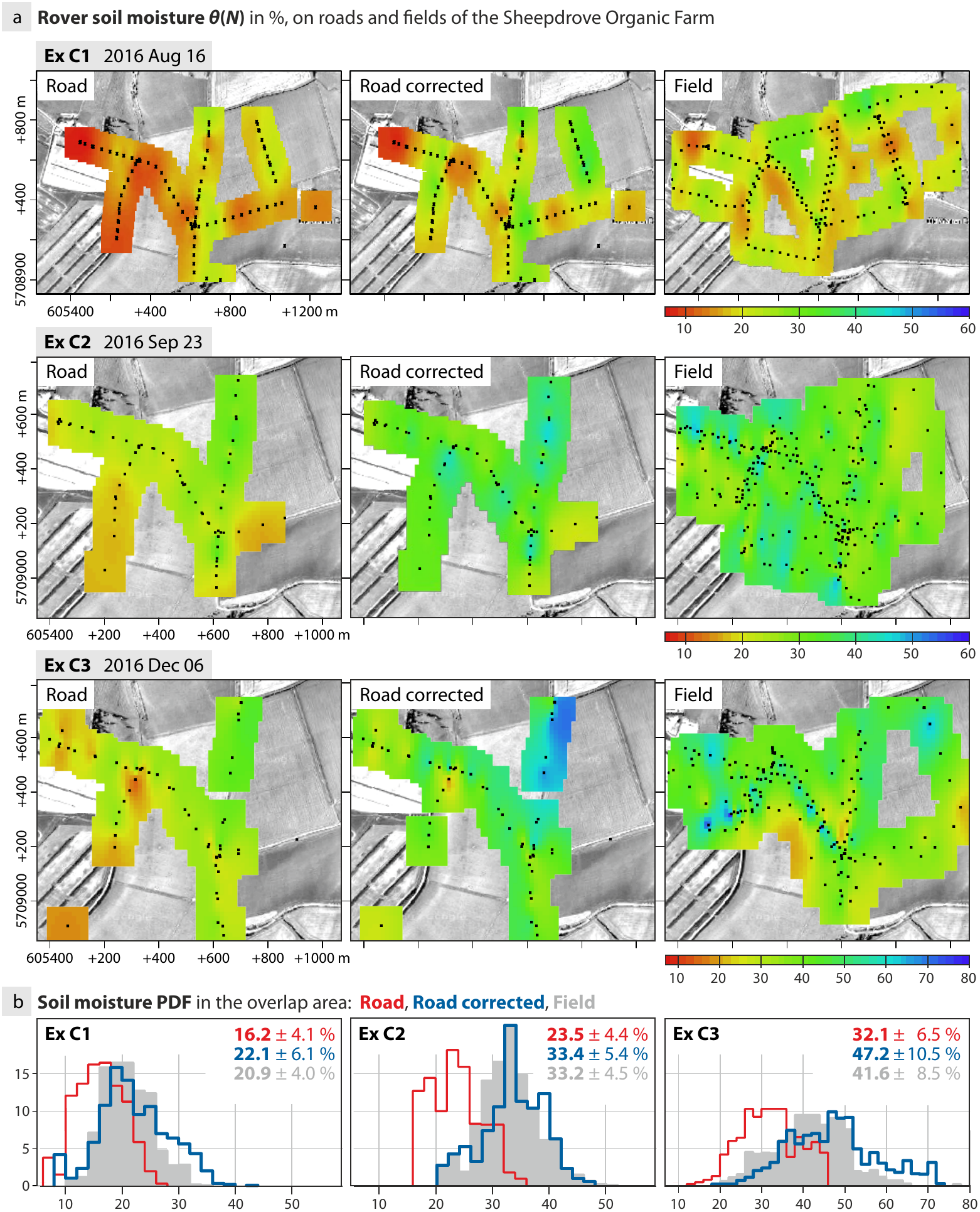}
\caption{(a) Interpolated soil moisture inferred from rover measurements for experiments C1, C2, C3, showing uncorrected road data (left), corrected road data (middle), and field data (right). (b) Probability density functions of data in the overlapping areas show that the corrected road data (blue) can represent the field data (grey) better than the uncorrected road data (red). Mean and standard deviation of each distribution are provided in the top right corner.}\label{fig:sfresults}
\end{figure}

On three different campaign days, the roads and the surrounding fields in the central valley of the \emph{Sheepdrove Farm} were surveyed with the CRNS rover. Road points have been corrected using eqs. \ref{eq:croad} and \ref{eq:croad2} based on the road types shown in Fig.~\ref{fig:maps}. The corresponding soil moisture maps and histograms (PDFs) are shown in Fig.~\ref{fig:sfresults}a and \ref{fig:sfresults}b, respectively, where the three campaign days are denoted as C1, C2, and C3.

In experiment C1, it was only possible to access the borders of the field (\(r=10\pm5\,\)m) due to farming activities. Nevertheless, the correction of the road dataset led to adequate improvement of the average soil moisture distribution (Fig.~\ref{fig:sfresults}b). However, some patterns were not adequately resolved by the road survey. According to the field measurements, the central northern field is wetter than the central southern field. From measurements on the road, only an average water content is seen with no distinction of the two fields. There are also discrepancies in the eastern part of the farm, where road and field patterns seem to be inverse. It can be speculated that one reason for this behavior is the influence of the south-eastern field, which has not been surveyed on that day. The dry spot at the north-west corner is due to buildings and a large concrete area, which were not accounted for in the correction procedure.

In experiment C2 it was possible to fully cross the fields to generate an adequate interpolation of field soil moisture. The correction of all road types appeared to agree very well with the overall pattern of the field measurements. The probability density functions show good agreement in the overlapping area of both datasets (i.e., near the road).
The road correction in Experiment C3 is also able to capture the patterns seen by the field survey, with the exception of the wet region in the northern part. It is speculative whether local ponds on the road, soaked soil, or local vegetation is influencing the data seen by the rover. This pathway is lowered by 1--2 meters compared to the field, and it is yet unknown whether small-scale heterogeneity of terrain features close to the sensor influences the CRNS performance. Additional vegetation correction could probably reduce the apparent soil moisture in this part, which is surrounded by unmanaged grass and hedges. In any case, low precipitation (drizzle) might have added interception water during this day, which is almost impossible to quantify.

All in all, the mean and standard deviation of soil moisture patterns could be restored by the application of the correction function on road points. Some of the field patterns were invisible from the road, especially when the fields on the left and on the right exhibit different water content, or when local hedges next to the roads contain or intercept water that shields the signal from the field.

\subsection{Tradeoff between measurements from the road and the field}\label{tradeoff-between-measurements-from-the-road-and-the-field}

Although it is evident that neutron measurements on the road can be biased substantially, it remains a challenge for experimentalists to access non-road areas, while either the access of fields is restricted or campaigns are required to cover large areas in a reasonable amount of time. Hence, roving on roads is much more practical and a necessary condition to travel from site to site. The campaigns in the \emph{Sheepdrove Farm} combine both, road and field data, from which it could be inferred which number of measurements in the field is needed, in addition to the road data, to obtain an acceptable estimate of the field soil moisture.

\begin{figure}[h]
\centering
\includegraphics{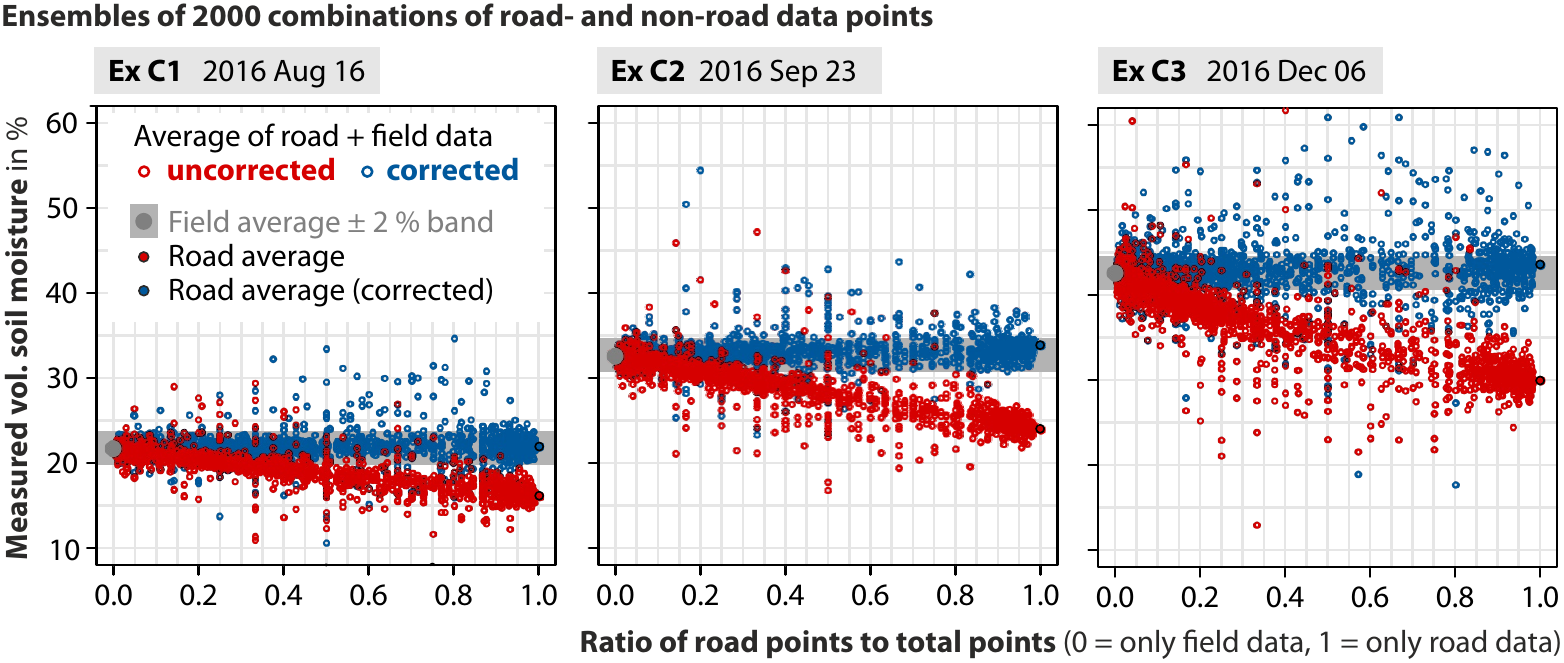}
\caption{Apparent average soil moisture of 2000 combinations of road- and non-road (i.e., field) points in experiments C1--3 plotted over the fraction of selected road points in the ensemble to selected total points (i.e., sum of road and field). Data is shown for uncorrected (red) and corrected (blue) road effect. If data is not corrected, a maximum fraction of road points between 0.20 and 0.05 is acceptable (under dry or wet conditions, respectively) to provide a realistic estimate of field soil moisture within \(2\,\%\) accuracy.}\label{fig:rafaelsplot}
\end{figure}

In Fig.~\ref{fig:rafaelsplot} all data points obtained during each campaign were bootstrapped, leading to more than 2000 combinations of road and field measurements.
The figure shows that the inclusion of uncorrected road points (red) can lead to an unreliable average value. Depending on wetness conditions, at least 80--\(95\,\%\) of the data points should be taken in the field to obtain an average that is within a two percent accuracy range around the mean field water content. In contrast to uncorrected data, corrected road data (blue) is already a good predictor for field soil moisture when any number of survey points on the road and in the field were averaged.

The analysis shows that any combination of field data and corrected road data can lead to a sufficient estimation of average water content in the survey area. However, the correction procedure is highly sensitive to supporting information like road moisture, field moisture, and road width (see Fig.~\ref{fig:croadsm}). If these parameters are uncertain, their impact on the CRNS product could be substantial. The impact could be reduced by calibrating the road-correction parameters with road and field data at selected anchor locations, or by including a substantial number of field data points to the dataset measured only on roads.

\section{Conclusions}\label{conclusions}

The mobile cosmic-ray neutron sensor (CRNS rover) was successfully applied to estimate soil moisture at scales from a few meters to a few square kilometers. One of the most prominent insights from the detailed and extensive investigations is the confirmation that the CRNS rover is capable to capture small-scale patterns at resolutions of 10--100 m, depending on driving speed. This result opens the path for non-invasive tomography of root-zone soil moisture patterns in small catchments and agricultural fields, where traditional methods would require exhaustive and time-consuming efforts.

On the other hand, the study revealed the critical need to apply correction approaches to account for local effects of dry roads. The different experiments carried out in the course of this study showed a critical loss in the capability to estimate average field soil moisture when the field was not accessible and the measurements were taken on roads only. This effect has been quantified in this study for the first time using neutron transport simulations, and has been confirmed by dedicated experiments.

We propose an analytical correction function which accounts for various road types and soil moisture conditions. As the analytical form of the corresponding relations has been based on physical reasoning, and the parameters were determined with the help of neutron simulations, the approach could be assumed to be not site-specific and universally applicable. However, the analytical fit showed a few limitations for roads that are wetter than the field, and for road widths beyond \(7\,\)m. The approach is further sensitive to the road parameters like width \(w\) and moisture \(\theta_\text{road}\). While the measurement of road moisture content can be impractical, this quantity could be treated as a calibration parameter by comparing data on the road and in the field at certain anchor locations.

The presented correction approach further depends on prior knowledge of field soil moisture (eqns.~\ref{eq:croad}, \ref{eq:croadp}). To circumvent this requirement, an adaption of the equation has been proposed that takes the uncorrected first-order approximation, \(\theta(N)\), as a proxy instead (eq.~\ref{eq:croad2}). Although we have shown its performance for the two, climatologically similar sites on ten different days throughout the years, the empirical character of this alternative approach requires more test cases under more various sites and conditions.

The corrected road data has been compared with field soil moisture inferred from independent TDR (experiment A) and rover measurements (experiments B and C). In all cases the corrected soil moisture product sensed from the road showed remarkable agreement with the patterns, the mean, and the standard deviation of soil moisture in the field. However, a few limitations have been identified. If strong differences in soil moisture are present between neighboring fields passed by the rover, it may be not possible for the sensor to capture the corresponding patterns due to the isotropic nature of neutron detection. Moreover, local ponds on pathways or nearby unmanaged vegetation could further influence the neutron signal in a way that is not representative for the field behind.

Nevertheless, a considerable amount of uncertainty is introduced to measurements from roads due to the high contribution of non-field neutrons and the uncertain properties of the road and its surroundings. Therefore, it is advisable to drive directly on the field wherever possible, or to take additional measurements on the field every now and then. This is advisable not only to make sure that the parameters of the road correction lead to a proper representation of field soil moisture, but also to support spatial interpolation.

Based on the conclusions above, we generally recommend to correct for the road effect before spatial CRNS data is used to support hydrological models or agricultural decisions. With regards to evaluation of remote-sensing products \citep[e.g.,][]{Chrisman2013} dry roads are also part of the remotely sensed average soil moisture, so that different correction approaches might be needed to compare both area-averaged products. There might also be ways to reduce the contribution of the roads in future developments of the neutron detector. For example, by mounting the detector on top of the car where it is more exposed to far-field neutrons.



%
%
%

%
%

%

%

%

\acknowledgments
Data is available from the authors on request. MS, RR, and JI thank Dan Bull for providing access to the Sheepdrove Organic Farm. MS, IS, and UW thank Thomas Grau, Mandy Kasner, and Andreas Schmidt for their support during field campaigns in the Sch{\"a}fertal. MS acknowledges kind support by the Helmholtz Impulse and Networking Fund through Helmholtz Interdisciplinary School for Environmental Research (HIGRADE). JI is funded by the Queen's School of Engineering, University of Bristol, EPSRC, grant code: EP/L504919/1. RR, JI and Sheepdrove Organic Farm activities are funded by the Natural Environment Research Council (A MUlti-scale Soil moistureEvapotranspiration Dynamics study (AMUSED); grant number NE/M003086/1). The research was funded and supported by the Terrestrial Environmental Observatories (TERENO), which is a joint collaboration program involving several Helmholtz Research Centers in Germany.

%
%
%
%
%
%
%
%
%







\end{document}